\documentstyle[prd,aps,preprint,tighten,epsf,amstex]{revtex}

\def\cancel#1#2{\ooalign{$\hfil#1\mkern1mu/\hfil$\crcr$#1#2$}}
\def\slash#1{\mathpalette\cancel{#1}}

\def\mpi2{m_\pi^2}
\def\mK2{m_K^2}

\newcommand{\bea}{\begin{eqnarray}}
\newcommand{\eea}{\end{eqnarray}}
\newcommand{\be}{\begin{equation}}
\newcommand{\ee}{\end{equation}}

\begin{document}
\bibliographystyle{apsrev}
\epsfclipon


\newcounter{Outline}
\setcounter{Outline}{1}

\newcounter{Intro}
\setcounter{Intro}{1}

\newcounter{Actions}
\setcounter{Actions}{1}

\newcounter{Simulation}
\setcounter{Simulation}{1}

\newcounter{Results}
\setcounter{Results}{1}

\newcounter{Conclusions}
\setcounter{Conclusions}{1}

\newcounter{Acknowledgments}
\setcounter{Acknowledgments}{1}

\newcounter{Appendix}
\setcounter{Appendix}{0}

\newcounter{Tables}
\setcounter{Tables}{1}

\newcounter{Figures}
\setcounter{Figures}{1}


\draft

\preprint{CU-TP-1169, BNL-NT-06/48}

\title{Flavor symmetry breaking and scaling for improved staggered actions in quenched QCD}

\author{
M.~Cheng$^a$,
N.~Christ$^a$,
C.~Jung$^b$,
F.~Karsch$^{b,d}$,
R.~Mawhinney$^a$,
P.~Petreczky$^{b,c}$,
K.~Petrov$^e$}

\address{
\vspace{0.5in}
$^a$Physics Department,
Columbia University,
New York, NY 10027}

\address{$^b$Physics Department,
Brookhaven National Laboratory,
Upton, NY 11973}

\address{$^c$RIKEN-BNL Research Center,
Brookhaven National Laboratory,
Upton, NY 11973}

\address{$^d$Fakult\"at f\"ur Physik,
Universit\"at Bielefeld,
D-33615 Bielefeld, Germany}

\address{$^e$Niels Bohr Institute,
University of Copenhagen,
Blegdamsvej 17, DK-2100 Copenhagen, Denmark}

\date{\today}
\maketitle

\begin{abstract}
We present a study of the flavor symmetry breaking in the pion spectrum for
various improved staggered fermion actions.  To study the effects of
link fattening and tadpole improvement, we use three different variants of
the p4 action - p4fat3, p4fat7, and p4fat7tad.  These are compared to Asqtad
and also to naive staggered.  To study the pattern of symmetry breaking, we
measure all 15 meson masses in the 4-flavor staggered theory.  The measurements
are done on a quenched gauge background, generated using a one-loop improved
Symanzik action with $\beta=10/g^2 = 7.40, 7.75,$ and $8.00$, corresponding
to lattice spacings of approximately a = .31 fm., .21 fm., and .14 fm.  We
also study how the lattice scale set by the $\rho$ mass on each of these
ensembles compares to one set by the static quark potential.
\end{abstract}


\newpage


\section{Introduction}
\label{sec:intro}

\ifnum\theIntro=1
%
%

Lattice QCD is the only first-principles technique to calculate various
important quantities where the strong nuclear force is an important effect.
However, the discretization of QCD on a lattice is plagued by many
difficulties.  Among them is the existence of doubler modes, which turns
one flavor of quark in the continuum limit into 16 flavors on a lattice.
Several different methods have been used to eliminate these doubler
modes.

One such method was introduced by Kogut and Susskind and
involves a thinning of the fermion degrees of freedom by ``staggering'' 
the components of the Dirac spinor over a four-dimensional hypercube.  
While this only reduces the number of doublers from 16 to 4, staggered
fermions are useful in that a remnant $U(1)$ chiral symmetry is preserved.

However, staggered fermions mix the spin
and flavor degrees of freedom, breaking flavor symmetry.  As a 
result, of the 15 pions in the 4 flavor staggered theory, only the
lightest pion is a true Goldstone boson, tending to $m_\pi \rightarrow 0$ as 
$m_q \rightarrow 0$.  The other fourteen pions have their masses increased
by $O(a^2)$ terms in the lattice action that do not
preserve flavor symmetry.  These mass splittings fall into a pattern that 
reflects the remnant chiral symmetry on the lattice.

Several attempts have been made to improve the
staggered fermion action, either by adding new terms to the fermion action 
(Naik
\cite{Naik:1986bn}, p4\cite{Heller:1999xz}) or by using some variant of 
gauge-link smearing\cite{Orginos:1999cr}\cite{Blum:1996uf}.  In this work, we examine
how well some of these improvements do in reducing flavor-symmetry breaking
by calculating the masses of all 15 pions in the staggered theory on a
quenched background.  The flavor symmetry of the pion spectrum has been
previously examined on quenched lattices with the naive staggered action
\cite{Ishizuka:1993mt} and on dyanmical lattices with the Asqtad action
\cite{Aubin:2004wf}.

The study of flavor symmetry breaking is especially important for simulations
of finite-temperature QCD, where chiral symmetry and the dynamics of
the the lightest mesons play a crucial role in determining the universal
properties of the QCD phase transition.  
Different variants of the staggered fermion action
are used in finite-temperatures simulations because they are numerically
inexpensive and naturally allow for $O(a^2)$ improvement through the
addition of a Naik or p4 term\cite{Karsch:2000ps}\cite{Bernard:2004je}.  
Since finite-temperature simulations
are currently only feasible for rather small temporal extents 
($N_\tau = 4,6,8$), and thus very coarse lattices, flavor symmetry 
improvement is vital for this purpose.

In addition, we examine the scaling properties of 
these actions by comparing the lattice scales determined by $m_\rho$ to
that set by the static quark potential ($r_1$).  
Typically, for the naive staggered action,
the ratio of the scales set by $m_\rho$ and $r_1$ depends strongly on the 
lattice spacing,
but studies\cite{Bernard:1999xx} have shown that the Asqtad action has
much better scaling properties than naive the staggered action.  Here we 
test the scaling properties of the various p4 actions described above.

\fi


\section{Actions}
\label{sec:actions}

\ifnum\theActions=1
%
%


\subsection{Staggered Action}
The Kogut-Susskind action is written as:
\begin{equation}
S_F = \sum_x \sum_{\mu}\eta_\mu(x)\left[\bar{\chi}(x)(U_\mu(x)\chi(x+\mu) - U^{\dagger}_\mu(x-\mu)\chi(x-\mu))\right] + 2m \sum_x \bar{\chi}(x)\chi(x)
\end{equation}
where $\chi(x)$ is a one-component spinors and $\eta_\mu(x) = (-1)^{x_0+...x_{\mu-1}}$ are the staggered phases.

By distributing the spinor degrees of freedom over a hypercube, the spin
and flavor degrees of freedom are mixed by $O(a^2)$ terms in the action.  
As outlined by Golterman and Smit\cite{Golterman:1985dz},
there are multiple meson states on the lattice that correspond to distinct
physical states.  For example, there are 15 different operators, falling into
7 distinct irreducible representations that have the quantum numbers of a
physical pion.

Because only a $U(1)$ remanant of the continuum chiral symmetry group is
preserved, only one of these 15 pions is a true Goldstone boson.  The
other 14 have masses have non-vanishing masses that are determined by
the $O(a^2)$ terms in the action.

\subsection{Improved Staggered Actions}
There have been several variants of staggered fermions that seek to improve
upon the naive staggered formulation.  Two notable examples that are the p4
\cite{Heller:1999xz} and Asqtad\cite{Orginos:1999cr} actions.  
Both of these actions incorporate an additional three-link term in the 
fermion derivative.  The p4 action includes a bent ``knight's move'' term,
while Asqtad uses the straight Naik term\cite{Naik:1986bn}.  These two
terms are the minimal allowed modifications of the fermion derivative
consistent with the symmetries of the staggered formulation.

A general action with both the Naik and the knight's move term is written as:

\begin{eqnarray}
  S_F & = &       2m\sum_{x}\bar{\chi}(x)\chi(x) + 
  \sum_{x}\bar{\chi}(x)\sum_{\mu}\eta_{\mu}(x)\nonumber\\
       & &     \Big\{ c_{1,0} \Big[U_{\mu}(x)\chi(x+\mu) - U_{\mu}^{\dagger}
	      (x-\mu) \chi(x-\mu)\Big] + {} \\
       & &  + c_{3,0}\Big[U_{\mu}^{(3,0)}(x)\chi(x+3\mu)-U_{\mu}^{(3,0)\dagger}
            (x-3\mu)\chi(x-3\mu)\Big] + {} \nonumber \\
      & & + c_{1,2}\sum_{\nu\neq\mu}\Big[U_{\mu,\nu}^{(1,2)}(x)
               \chi(x+\mu+2\nu)- U_{\mu,\nu}^{(1,2)\dagger}(x-\mu-2\nu) 
               \chi(x-\mu-2\nu)+ {}\nonumber\\
      & & \qquad \qquad + U_{\mu,\nu}^{(1,-2)}(x)\chi(x+\mu-2\nu)
	  -U_{\mu,\nu}^{(1,-2)\dagger}(x-\mu+2\nu)\chi
          (x-\mu-2\nu)\Big] \Big\} \nonumber
\end{eqnarray}
where
\begin{eqnarray}
U_{\mu}^{(3,0)}(x) & = & U_{\mu}(x)U_{\mu}(x+\mu)U_{\mu}(x+2\mu) \\
U_{\mu,\nu}^{(1,2)}(x) & = & \frac{1}{2}\big[ U_{\mu}(x)U_{\nu}(x+\mu)
U_{\nu}(x+\mu+\nu)+U_{\nu}(x)U_{\nu}(x+\nu)U_{\mu}(x+2\nu)\big] \nonumber \\
U_{\mu,\nu}^{(1,-2)}(x) & = & \frac{1}{2}\big[ U_{\mu}(x)U_{\nu}^{\dagger}
(x+\mu-\nu)U_{\nu}^{\dagger}(x+\mu-2\nu)+U_{\nu}^{\dagger}(x-\nu)
U_{\nu}^{\dagger}(x-2\nu)U_{\mu}(x-2\nu)\big] \nonumber
\end{eqnarray}

For the p4 action, $c_{1,2} = 1/24, c_{3,0} = 0$; the Naik action
has $c_{1,2} = 0, c_{3,0} = -1/48$.  In the free-field limit, the Naik action
eliminates the $O(a^2)$ errors in the quark propagator.  The p4 action is 
chosen to remove the violations to rotational symmetry in the quark
propogator through $O(p^4)$.

\subsection{Gauge Link Smearing}
It has beens shown that gauge link smearing helps reduce the amount of flavor
symmetry violation in the fermion action\cite{Orginos:1999cr}.  
Gauge link smearing involves
replacing the one-link term in the fermion action with some linear combination
of gauge invariant staples that connect nearest-neighbor sites.
\vspace{-12mm}
\begin{center}
\begin{equation}
\epsfxsize = 0.55\textwidth
\epsfbox{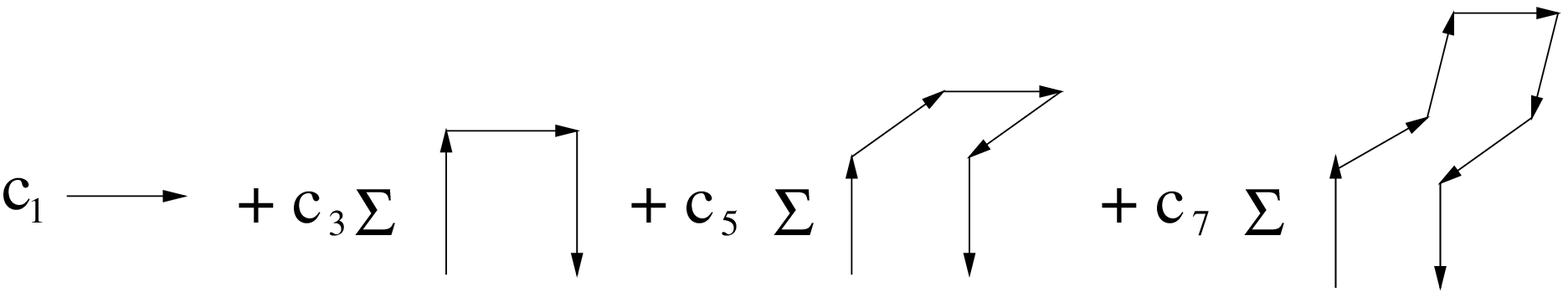}
\end{equation}
\end{center} 

The p4fat3 action has been used previously in thermodynamic simulations
\cite{Karsch:2000kv}.  Only the three-link staple is included for p4fat3.
The smearing coefficients for the Asqtad action are chosen to cancel 
out the tree-level violations of flavor symmetry.  
The Asqtad action is also tadpole-improved, and includes the planar, 
5-link Lepage term to cancel out the $O(a^2)$ errors introduced by 
the smearing procedure\cite{Orginos:1999cr}.  The smearing 
coefficients for p4fat7 and p4fat7tad are determined in the same manner
as in Asqtad, but the Lepage term is omitted.  In addition, we compare the
tadpole-improved p4fat7tad action with the p4fat7 action in order to test the 
effects of tadpole improvement
A full summary of the smearing parameters can be found in 
Table \ref{tab:actions}.

\label{subsec:}

\fi


\section{Simulation}
\label{sec:simulation}

\ifnum\theSimulation=1
%
%

\subsection{Gauge Action}
To test the properties of the various actions, we measure the meson
spectrum on  quenched gauge backgrounds with $\beta = 10/g^2 =  7.40, 7.75, 
8.00$ using a one-loop improved Symanzik action.  For each value of $\beta$, 
we generated 100 configurations.  For $\beta = 7.40, 
7.75$, we used a volume of $16^3\times 32$ and separated configurations by 
1000 sweeps of the gauge heat bath algorithm.  For the finest ensemble, 
$\beta = 8.00$, we separated configurations by 500 sweeps, on a volume of 
$24^3\times 32$.  These gauge configurations match those used previously to 
test the scaling behavior of the Asqtad action\cite{Bernard:1999xx}.

\subsection{Sources}
Following Ref. \cite{Ishizuka:1993mt}, we calculate the quark propagator for eight different wall sources
on each configuration by solving using the standard conjugate gradient technique.
\begin{eqnarray}
\sum_{x'}(D D^\dagger)(x,x') ~X^{\vec{A}}(x') & = & \sum_{2\vec{y}}\delta_{\vec{x},2\vec{y}+\vec{A}} ~\delta_{x_4,0}\\
\chi^{\vec{A}}(x) & = & \sum_{x'} D^{\dagger}(x,x')~X^{\vec{A}}(x')
\end{eqnarray}
where the summation over $2\vec{y}$ goes over all the unit cubes at time slice $t = 0$ and $\vec{A}$ labels each of the eight different corners of a 3-D unit cube.  We combine the quark propagators that we obtain into meson propagators:
\begin{eqnarray}
M_{\vec{A},\vec{B}} & = & \sum_x D_{\vec{A}}~\bar{\chi}^{\vec{A}}(x)~D_{\vec{B}}~\chi^{\vec{B}}(x)\\
D_{\vec{A}} \chi(x) & = & \sum_{\delta_1=\pm 1}\sum_{\delta_2=\pm 1}\sum_{\delta_3=\pm 1} \chi(x + \delta_1 A_1 {\bf e_1}+ \delta_2 A_2 {\bf e_2} + \delta_3 A_3 {\bf e_3})
\end{eqnarray}
The 64 different meson operators fall into irreducible representations that can be written as linear combinations of the meson propagators defined above using the symmetric shift operator $D_{\vec{A}}$.
\begin{equation}
M(t) = \sum_{\vec{A}} \phi(\vec{A})M_{\vec{A},(\vec{A}+\vec{\delta})}
\end{equation}
$\vec{\delta}$ and $\phi(\vec{A})$ are specified for each irreducible 
representation.  Table \ref{tab:mesons} lists the details 
for all the meson
representations.  For us, only the meson operators that correspond to a 
physical pion are of interest.  Henceforth, we shall label the meson states
using the conventions set forth in \cite{Ishizuka:1993mt}, which labels
meson states by their transformation properties under the lattice symmetry
group ($\bold{r}^{\sigma_s \sigma_{123}}$). 

\subsection{Measurements}
For each set of configurations, we measured the 15 different
$\pi$ propagators for at least three different values of $m_q$.  
These values are chosen so that $m_\pi$ falls into approximately
the same range for the different actions.  By using three masses,
we can extrapolate to the $m_q\rightarrow 0$ limit.  Values of $m_{\pi}$
were then extracted by fitting to the form:
\begin{equation}
M(t) = A \left(e^{-m_+ t} + e^{-m_+ (T-t)} \right) + B~ (-1)^t \left(e^{-m_- t} + e^{-m_- (T-t)} \right)
\end{equation}
where $m_+$ and $m_-$ denote the masses of opposite parity states.

For $\beta = 8.00$, we use a fitting range of $6-16$; for $\beta = 7.75$, 
we use a range of $5-16$; and for $\beta = 8.00$, we use $4-16$ except for
staggered fermions, where the larger masses requires us to use a fitting
range of $3-16$ to obtain a good signal.

Tables \ref{tab:asqtad}-\ref{tab:staggered}
show the values for $m_\pi$ that we have obtained.  
While we measure masses for all 15 operators, several of the operators 
are related by lattice
symmetries and fall into degenerate triplets.  Our data confirms this
expected degeneracy, so we have quoted only one value for each of these
triplets by averaging together the propagators for each of the operators.
As a result, we are left with 7 distinct masses for each set of measurements.  

The errors are calculated by using the jackknife method with a block size of
one configuration.  While we only quote statistical errors, there is an
additional systematic error associated with varying the fit range.
For most of the measurements, this effect is smaller than the quoted
statistical error.  However, for the coarsest lattices at $\beta = 7.40$, 
some of the operators have a poorer signal and the resulting masses
have systematic errors that may exceed the statistical errors in some cases.

The vector meson mass $m_\rho$ is measured using the same procedure as
outlined above, but we measure only the local representation 
$\bold{r}^{\sigma_s \sigma_{123}}= 3''++$.  The fitting ranges that
we use for $m_\rho$ match those used for the pions.

\fi


\section{Results}
\label{sec:results}

\ifnum\theResults=1
%
%


\subsection{Flavor symmetry breaking}
Our data in Tables \ref{tab:asqtad}-\ref{tab:staggered} show
that the $\pi$ spectrum for each fermion action falls into the same 
general pattern shown by previous studies\cite{Ishizuka:1993mt}.  
The state corresponding to the Goldstone pion  ($r^{\sigma_s \sigma_{123}} = 
1+-$) is clearly lighter than the other pions.  We also expect
the splittings to be $O(a^2)$, and indeed the mass differences
are smallest for the finest lattice ($\beta = 8.00$) and largest for the
coarsest lattice ($\beta = 7.40$).  Because we use a Symanzik-improved 
gauge action, the mass splittings for all actions are also reduced
compared to unimproved gauge actions\cite{DeGrand:2002vu}.

In general, we expect the different representations to yield different
masses.  However some of the representations  are degenerate at $O(a^2)$, 
differing only at $O(a^4)$\cite{Lee:1999zx}.  In particular, we expect 
the following pairs of representations to be nearly degenerate: $1++$ 
and $3''--$, $3''-+$ and $3''+-$, and $3''++$ and $1--$. 
This approximate degeneracy is evident for each of the fermion actions.
Our precision is not enough to discern the expected pattern of $O(a^4)$ 
splittings.

Using the data collected in Tables \ref{tab:asqtad}-\ref{tab:staggered}, we
extrapolate each of the 15 different pions to the chiral limit 
($m_q \rightarrow 0$).  These fits were done using the expected 
chiral form:
\begin{equation}
m_\pi^2(m_q) = m_\pi^2(m_q=0) + B m_q
\end{equation}
Figure \ref{fig:chiral_extr} shows the result of a typical chiral fit for
each of the 7 distinct representations.  
Tables \ref{tab:chiral740}-\ref{tab:chiral800} give the full result of
these extrapolations.  In the continuum limit, we expect the Goldstone pion 
to have vanishing mass as $m_q \rightarrow 0$. In our case, $m_\pi$ does not
vanish at $m_q=0$ due to finite volume effects. 

Figures \ref{fig:chiral740}-\ref{fig:chiral800} show the chiral extrapolated
values of $m_\pi$ for the different fermion actions.  The flavor symmetry
breaking, characterized by the mass of the non-Goldstone pions, is worst for
naive staggered fermions of all the actions tested.  The p4fat3 action, 
which uses three-link smearing, is a significant improvement over the 
naive staggered action.
However, the three highly-improved actions (Asqtad, p4fat7, p4fat7tad) show 
the best flavor-symmetry characteristics.  These three actions give
similar results.  For $\beta = 7.40$, p4fat7 and Asqtad agree to within 
statistical errors, while p4fat7tad is slightly better.  For $\beta = 7.75$
and $\beta = 8.00$, the p4 actions are slightly better than Asqtad, and
tadpole improvement seems to have little effect - p4fat7 and p4fat7tad
agree to within errors.  Figure \ref{fig:splitting} shows the mass splitting
as a function of $a^2$.  As expected, the quantity $m_\pi^2 - m_G^2$ vanishes
linearly with $a^2$ for each of the actions tested.

\subsection{$m_\rho$ vs. $r_1$ scaling}
We also measured the vector meson mass $m_\rho$ for several different values
of $m_q$.  Table \ref{tab:rho} gives $m_\rho$ extrapolated to the 
$m_q \rightarrow 0$ limit using the linear chiral form:
\begin{equation}
m_\rho(m_q) =  m_\rho(m_q=0) + B m_q
\end{equation}
We wish to compare the scale determined by $m_\rho$ with the parameter
$r_1$ extracted from the static quark potential.
The values for $r_1$ that we use are calculated
by the MILC collaboration on these lattices and are used in Ref. 
\cite{Bernard:1999xx} to check the scaling of the Asqtad action.  $r_1/a = 
1.44(1), 2.08(5), 2.653(10)$ for $\beta = 7.40, 7.75, 8.00$ respectively.

Figure \ref{fig:rhoscaling} shows $m_\rho r_1$ plotted against the lattice
spacing $(a/r_1)^2$.  For Asqtad and p4fat3 actions, $m_\rho r_1$ does
not change appreciably over this range.  For p4fat7 and p4fat7tad, we seem
to see a 10\% decrease in $m_\rho r_1$ from the finest lattice to the coarsest
lattice.
\fi


\section{Conclusions}
\label{sec:conclusions}

\ifnum\theConclusions=1
%
%

We have made a detailed comparison of the various different variants of
improved staggered actions as they relate to flavor symmetry breaking in the 
$\pi$ spectrum and the scaling of the $\rho$ mass.

As expected, the flavor symmetry violations are most severe for the coarsest
lattices.  All
four variants of improved staggered fermions (Asqtad, p4fat3, p4fat7, 
p4fat7tad) exhibit much better flavor symmetry properties than naive
staggered fermions.  It seems that gauge-link fattening is the most
important factor in these improvements.  While the p4fat3 action is
better than normal staggered, it is noticeably worse than the more
highly improved actions.  Among the three
actions which employ the most smearing (Asqtad, p4fat7, p4fat7tad), it seems
like the p4 variants are very slightly preferred to Asqtad.  Also, tadpole
improvement seems to have very little affect on the pattern of flavor
symmetry breaking.

As for $m_\rho$, we find that all the improved actions show only mild
deviations from scaling in the set of ensembles that we have examined.  All
of them perform better than naive staggered, although p4fat7 and p4fat7tad
show a 10\% variation in $m_\rho r_1$.

\fi


\bibliography{paper}

\begin{thebibliography}{10}
\expandafter\ifx\csname bibnamefont\endcsname\relax
  \def\bibnamefont#1{#1}\fi
\expandafter\ifx\csname bibfnamefont\endcsname\relax
  \def\bibfnamefont#1{#1}\fi
\expandafter\ifx\csname url\endcsname\relax
  \def\url#1{\texttt{#1}}\fi
\expandafter\ifx\csname urlprefix\endcsname\relax\def\urlprefix{URL }\fi
\expandafter\ifx\csname bibinfo\endcsname\relax \def\bibinfo#1#2{#2}\fi
\expandafter\ifx\csname eprint\endcsname\relax \def\eprint#1{#1}\fi

\bibitem{Naik:1986bn}
\bibinfo{author}{\bibfnamefont{S.}~\bibnamefont{Naik}}, \bibinfo{journal}{Nucl.
  Phys.} \textbf{\bibinfo{volume}{B316}}, \bibinfo{pages}{238}
  (\bibinfo{year}{1989}).

\bibitem{Heller:1999xz}
\bibinfo{author}{\bibfnamefont{U.~M.} \bibnamefont{Heller}},
  \bibinfo{author}{\bibfnamefont{F.}~\bibnamefont{Karsch}}, \bibnamefont{and}
  \bibinfo{author}{\bibfnamefont{B.}~\bibnamefont{Sturm}},
  \bibinfo{journal}{Phys. Rev.} \textbf{\bibinfo{volume}{D60}},
  \bibinfo{pages}{114502} (\bibinfo{year}{1999}), \eprint{hep-lat/9901010}.

\bibitem{Orginos:1999cr}
\bibinfo{author}{\bibfnamefont{K.}~\bibnamefont{Orginos}},
  \bibinfo{author}{\bibfnamefont{D.}~\bibnamefont{Toussaint}},
  \bibnamefont{and} \bibinfo{author}{\bibfnamefont{R.~L.} \bibnamefont{Sugar}}
  (\bibinfo{collaboration}{MILC}), \bibinfo{journal}{Phys. Rev.}
  \textbf{\bibinfo{volume}{D60}}, \bibinfo{pages}{054503}
  (\bibinfo{year}{1999}), \eprint{hep-lat/9903032}.

\bibitem{Blum:1996uf}
\bibinfo{author}{\bibfnamefont{T.}~\bibnamefont{Blum}} \emph{et~al.},
  \bibinfo{journal}{Phys. Rev.} \textbf{\bibinfo{volume}{D55}},
  \bibinfo{pages}{1133} (\bibinfo{year}{1997}), \eprint{hep-lat/9609036}.

\bibitem{Ishizuka:1993mt}
\bibinfo{author}{\bibfnamefont{N.}~\bibnamefont{Ishizuka}},
  \bibinfo{author}{\bibfnamefont{M.}~\bibnamefont{Fukugita}},
  \bibinfo{author}{\bibfnamefont{H.}~\bibnamefont{Mino}},
  \bibinfo{author}{\bibfnamefont{M.}~\bibnamefont{Okawa}}, \bibnamefont{and}
  \bibinfo{author}{\bibfnamefont{A.}~\bibnamefont{Ukawa}},
  \bibinfo{journal}{Nucl. Phys.} \textbf{\bibinfo{volume}{B411}},
  \bibinfo{pages}{875} (\bibinfo{year}{1994}).

\bibitem{Aubin:2004wf}
\bibinfo{author}{\bibfnamefont{C.}~\bibnamefont{Aubin}} \emph{et~al.},
  \bibinfo{journal}{Phys. Rev.} \textbf{\bibinfo{volume}{D70}},
  \bibinfo{pages}{094505} (\bibinfo{year}{2004}), \eprint{hep-lat/0402030}.

\bibitem{Karsch:2000ps}
\bibinfo{author}{\bibfnamefont{F.}~\bibnamefont{Karsch}},
  \bibinfo{author}{\bibfnamefont{E.}~\bibnamefont{Laermann}}, \bibnamefont{and}
  \bibinfo{author}{\bibfnamefont{A.}~\bibnamefont{Peikert}},
  \bibinfo{journal}{Phys. Lett.} \textbf{\bibinfo{volume}{B478}},
  \bibinfo{pages}{447} (\bibinfo{year}{2000}), \eprint{hep-lat/0002003}.

\bibitem{Bernard:2004je}
\bibinfo{author}{\bibfnamefont{C.}~\bibnamefont{Bernard}} \emph{et~al.}
  (\bibinfo{collaboration}{MILC}), \bibinfo{journal}{Phys. Rev.}
  \textbf{\bibinfo{volume}{D71}}, \bibinfo{pages}{034504}
  (\bibinfo{year}{2005}), \eprint{hep-lat/0405029}.

\bibitem{Bernard:1999xx}
\bibinfo{author}{\bibfnamefont{C.~W.} \bibnamefont{Bernard}} \emph{et~al.}
  (\bibinfo{collaboration}{MILC}), \bibinfo{journal}{Phys. Rev.}
  \textbf{\bibinfo{volume}{D61}}, \bibinfo{pages}{111502}
  (\bibinfo{year}{2000}), \eprint{hep-lat/9912018}.

\bibitem{Golterman:1985dz}
\bibinfo{author}{\bibfnamefont{M.~F.~L.} \bibnamefont{Golterman}},
  \bibinfo{journal}{Nucl. Phys.} \textbf{\bibinfo{volume}{B273}},
  \bibinfo{pages}{663} (\bibinfo{year}{1986}).

\bibitem{Karsch:2000kv}
\bibinfo{author}{\bibfnamefont{F.}~\bibnamefont{Karsch}},
  \bibinfo{author}{\bibfnamefont{E.}~\bibnamefont{Laermann}}, \bibnamefont{and}
  \bibinfo{author}{\bibfnamefont{A.}~\bibnamefont{Peikert}},
  \bibinfo{journal}{Nucl. Phys.} \textbf{\bibinfo{volume}{B605}},
  \bibinfo{pages}{579} (\bibinfo{year}{2001}), \eprint{hep-lat/0012023}.

\bibitem{DeGrand:2002vu}
\bibinfo{author}{\bibfnamefont{T.~A.} \bibnamefont{DeGrand}},
  \bibinfo{author}{\bibfnamefont{A.}~\bibnamefont{Hasenfratz}},
  \bibnamefont{and} \bibinfo{author}{\bibfnamefont{T.~G.}
  \bibnamefont{Kovacs}}, \bibinfo{journal}{Phys. Rev.}
  \textbf{\bibinfo{volume}{D67}}, \bibinfo{pages}{054501}
  (\bibinfo{year}{2003}), \eprint{hep-lat/0211006}.

\bibitem{Lee:1999zx}
\bibinfo{author}{\bibfnamefont{W.-J.} \bibnamefont{Lee}} \bibnamefont{and}
  \bibinfo{author}{\bibfnamefont{S.~R.} \bibnamefont{Sharpe}},
  \bibinfo{journal}{Phys. Rev.} \textbf{\bibinfo{volume}{D60}},
  \bibinfo{pages}{114503} (\bibinfo{year}{1999}), \eprint{hep-lat/9905023}.

\end{thebibliography}


\ifnum\theAcknowledgments=1
\section*{Acknowledgments}

We thank RIKEN, Brookhaven National Laboratory and the U.S.  Department
of Energy for providing the facilities essential for the completion of
this work.

The numerical computations were done on the QCDOC computers, located at
Columbia University and Brookhaven National Lab.
\fi


\appendix

\ifnum\theAppendix=1
%
%

\section{Conventions for States and Operators}

\ifnum\theOutline=1
\framebox{Begin \ outline}
\begin{enumerate}
\item Details for meson states
\end{enumerate}
\framebox{End \ outline}
\fi

Comparing the Lagrangian of chiral perturbation theory described
previously with the Lagrangian of QCD, defines the
relationship between quantities expressed in terms of the self-adjoint
fields of chiral perturbation theory and the quark fields used in our
simulations.  Our notation follows \cite{Bernard:1989nb}.
We start with the Minkowski space QCD Lagrangian
\begin{equation}
  {\cal L}_{\rm QCD} = -\frac{1}{4} (F_{\mu \nu}^a)^2
    + \overline{\psi} \left( i\slash{D} - m \right) \psi
\end{equation}

\fi


\ifnum\theTables=1


\begin{table}
\begin{tabular}{lccccccc}
Action & $c_1$ & $c_{Naik}$ & $c_{knight}$ & $c_3$ & $c_5$ & $c_7$ & $c_{Lepage}$\\
\hline


Asqtad & $1/8+3/8+1/8$ & $-1/24u_0^2$ & $0$ & $1/16u_0^2$ & $1/64u_0^4$ & $1/384u_0^6$ & $-1/16u_0^4$\\
p4fat3& $15/44$ & $0$ & $1/24$ & $3/44$ & $0$ & $0$ & $0$\\
p4fat7 & $1/8-1/4$ & $0$ & $1/24$ & $1/16$ & $1/64$ & $1/384$ & $0$\\
p4fat7tad& $1/8-1/4$ & $0$ & $1/24u_0^2$ & $1/16u_0^2$ & $1/64u_0^4$ & $1/384u_0^6$ & $0$\\
staggered & $1$ & $0$  & $0$  & $0$  & $0$  & $0$  & $0$\\

\end{tabular}
\caption{Parameters for the different fermion actions. $u_0$ is the tadpole-improvement coefficient, $u_0 = \left<\Box\right>^{1/4}$}
\label{tab:actions}
\end{table}


\begin{table}
\begin{tabular}{lcc}
$r^{\sigma_s \sigma_{123}}$ & $\phi(\vec{A})$ & $\delta$\\
\hline


$1++$ & 1 & 0\\
$1+-$ & $\eta_4\zeta_4$ & 0\\
$3''--$& $\zeta_k\epsilon$ & $\hat{\bold{k}}$\\
$3''-+$& $\eta_4\zeta_4\zeta_k\epsilon$ & $\hat{\bold{k}}$\\
$3''++$& $\zeta_k\zeta_l\epsilon_{klm}$ & $\hat{\bold{k}}+\hat{\bold{l}}$\\
$3''+-$& $\eta_4\zeta_4\zeta_k\zeta_l\epsilon_{klm}$ & $\hat{\bold{k}}+\hat{\bold{l}}$\\
$1--$& $\eta_4\zeta_4\eta_1\eta_2\eta_3$ & $\hat{\bold{1}}+\hat{\bold{2}}+\hat{\bold{3}}$\\

\end{tabular}
\caption{Staggered Pions. $\eta_\mu = (-1)^{A_1+...A_{\mu-1}}, \zeta_\mu = (-1)^{A_{\mu+1}+...A_4}, \epsilon = (-1)^{A_1+...A_4}$, $\hat{\bold{k}} \neq \hat{\bold{l}}$}
\label{tab:mesons}
\end{table}    


\begin{table}
\begin{tabular}{l|r@{.}lr@{.}lr@{.}l|r@{.}lr@{.}lr@{.}l|r@{.}lr@{.}lr@{.}l}
$\bold{r}^{\sigma_s \sigma_{123}}$ & \multicolumn{6}{|c|}{$\beta=7.40$} & \multicolumn{6}{|c|}{$\beta=7.75$} & \multicolumn{6}{|c}{$\beta=8.00$}\\
\hline
$m_qa = $ & &02 & &03 & &04 & &02 & &03 & &04 & &01 & &02 & &03\\
\hline

$1+-$ & 0&3605(3) & 0&4390(3) & 0&5047(3) & 0&314(1) & 0&383(1) & 0&440(1) & 0&2003(5) & 0&2800(5) & 0&3409(5) \\
$1++$ & 0&644(8) & 0&690(4) & 0&734(3) & 0&399(2) & 0&455(2) & 0&504(1) & 0&247(1) & 0&3145(10) & 0&3693(8) \\
$3''--$ & 0&648(2) & 0&698(1) & 0&7444(9) & 0&401(2) & 0&457(1) & 0&506(1) & 0&2474(7) & 0&3142(7) & 0&3689(6) \\
$3''-+$ & 0&78(1) & 0&824(6) & 0&863(5) & 0&455(3) & 0&504(2) & 0&548(2) & 0&278(2) & 0&338(1) & 0&3883(10) \\
$3''++$ & 0&90(2) & 0&93(1) & 0&96(1) & 0&502(4) & 0&543(3) & 0&582(3) & 0&303(2) & 0&358(2) & 0&405(1) \\
$3''+-$ & 0&802(3) & 0&840(2) & 0&877(2) & 0&456(2) & 0&504(1) & 0&549(1) & 0&2791(8) & 0&3379(7) & 0&3882(7) \\
$1--$ & 0&90(1) & 0&935(8) & 0&968(6) & 0&498(3) & 0&541(2) & 0&582(2) & 0&306(1) & 0&3585(8) & 0&4049(8) \\

\end{tabular}
\caption{Asqtad pion masses}
\label{tab:asqtad}
\end{table}    


\begin{table}
\begin{tabular}{l|r@{.}lr@{.}lr@{.}l|r@{.}lr@{.}lr@{.}l|r@{.}lr@{.}lr@{.}l}
$\bold{r}^{\sigma_s \sigma_{123}}$ & \multicolumn{6}{|c|}{$\beta=7.40$} & \multicolumn{6}{|c|}{$\beta=7.75$} & \multicolumn{6}{|c}{$\beta=8.00$}\\
\hline
$m_qa = $ & &02 & &03 & &04 & &02 & &03 & &04 & &01 & &02 & &03\\
\hline

$1+-$ & 0&3136(2) & 0&4395(2) & 0&5346(2) & 0&2781(6) & 0&3893(9) & 0&474(1) & 0&2469(4) & 0&3459(5) & 0&4218(5) \\
$1++$ & 0&75(3) & 0&80(1) & 0&861(6) & 0&416(3) & 0&503(2) & 0&576(2) & 0&310(2) & 0&395(1) & 0&4650(8) \\
$3''--$ & 0&721(3) & 0&792(1) & 0&857(1) & 0&420(1) & 0&504(1) & 0&576(1) & 0&3094(7) & 0&3941(6) & 0&4638(6) \\
$3''-+$ & 0&94(4) & 0&96(2) & 1&012(10) & 0&495(4) & 0&569(3) & 0&634(2) & 0&348(2) & 0&425(1) & 0&4905(10) \\
$3''++$ & 1&05(8) & 1&11(4) & 1&14(2) & 0&549(6) & 0&614(4) & 0&673(3) & 0&378(3) & 0&448(1) & 0&510(1) \\
$3''+-$ & 0&904(8) & 0&966(3) & 1&023(3) & 0&495(2) & 0&568(2) & 0&633(2) & 0&3481(8) & 0&4244(7) & 0&4898(7) \\
$1--$ & 1&00(3) & 1&05(1) & 1&112(9) & 0&548(4) & 0&612(2) & 0&672(2) & 0&378(1) & 0&4477(9) & 0&5093(8) \\

\end{tabular}
\caption{p4fat3 pion masses}
\label{tab:p4fat3}
\end{table}    


\begin{table}
\begin{tabular}{l|r@{.}lr@{.}lr@{.}l|r@{.}lr@{.}lr@{.}l|r@{.}lr@{.}lr@{.}l}
$\bold{r}^{\sigma_s \sigma_{123}}$ & \multicolumn{6}{|c|}{$\beta=7.40$} & \multicolumn{6}{|c|}{$\beta=7.75$} & \multicolumn{6}{|c}{$\beta=8.00$}\\
\hline
$m_qa = $ & &02 & &03 & &04 & &02 & &03 & &04 & &01 & &02 & &03\\
\hline

$1+-$ & 0&4205(3) & 0&5832(3) & 0&7035(3) & 0&3216(8) & 0&449(1) & 0&5467(9) & 0&2727(6) & 0&3812(6) & 0&4651(6) \\
$1++$ & 0&681(6) & 0&798(3) & 0&896(2) & 0&395(2) & 0&504(2) & 0&593(1) & 0&302(1) & 0&4019(9) & 0&4821(7) \\
$3''--$ & 0&681(1) & 0&8003(9) & 0&8999(8) & 0&396(2) & 0&504(1) & 0&593(1) & 0&3021(7) & 0&4014(7) & 0&4815(6) \\
$3''-+$ & 0&789(6) & 0&895(4) & 0&982(3) & 0&442(3) & 0&539(2) & 0&621(2) & 0&323(2) & 0&416(1) & 0&4926(7) \\
$3''++$ & 0&90(1) & 0&978(6) & 1&054(4) & 0&484(4) & 0&569(3) & 0&645(2) & 0&341(2) & 0&428(1) & 0&5019(8) \\
$3''+-$ & 0&801(2) & 0&901(2) & 0&988(1) & 0&443(2) & 0&539(1) & 0&621(1) & 0&3233(8) & 0&4156(7) & 0&4924(7) \\
$1--$ & 0&894(6) & 0&980(4) & 1&057(2) & 0&483(2) & 0&569(2) & 0&645(2) & 0&3424(9) & 0&4283(8) & 0&5019(7) \\

\end{tabular}
\caption{p4fat7 pion masses}
\label{tab:p4fat7}
\end{table}    


\begin{table}
\begin{tabular}{l|r@{.}lr@{.}lr@{.}l|r@{.}lr@{.}lr@{.}l|r@{.}lr@{.}lr@{.}l}
$\bold{r}^{\sigma_s \sigma_{123}}$ & \multicolumn{6}{|c|}{$\beta=7.40$} & \multicolumn{6}{|c|}{$\beta=7.75$} & \multicolumn{6}{|c}{$\beta=8.00$}\\
\hline
 $m_qa = $& &02 & &03 & &04 & &02 & &03 & &04 & &01 & &02 & &03\\
\hline

$1+-$ & 0&4227(3) & 0&5139(3) & 0&5900(3) & 0&449(1) & 0&5467(9) & 0&6293(9) & 0&2161(6) & 0&3015(6) & 0&3669(6) \\
$1++$ & 0&643(4) & 0&708(3) & 0&766(2) & 0&504(2) & 0&593(1) & 0&671(1) & 0&250(2) & 0&326(1) & 0&3870(10) \\
$3''--$ & 0&6476(10) & 0&7124(9) & 0&7708(8) & 0&504(1) & 0&593(1) & 0&670(1) & 0&2518(9) & 0&3263(7) & 0&3866(7) \\
$3''-+$ & 0&762(5) & 0&816(4) & 0&867(3) & 0&539(2) & 0&621(2) & 0&695(1) & 0&276(2) & 0&345(2) & 0&402(1) \\
$3''++$ & 0&873(10) & 0&915(6) & 0&956(5) & 0&569(3) & 0&645(2) & 0&715(2) & 0&300(3) & 0&361(2) & 0&414(1) \\
$3''+-$ & 0&775(2) & 0&828(2) & 0&877(1) & 0&539(1) & 0&621(1) & 0&695(1) & 0&279(1) & 0&3452(8) & 0&4014(7) \\
$1--$ & 0&869(7) & 0&915(4) & 0&957(3) & 0&569(2) & 0&645(2) & 0&715(1) & 0&302(1) & 0&3622(8) & 0&4147(8) \\

\end{tabular}
\caption{p4fat7tad pion masses}
\label{tab:p4fat7tad}
\end{table}    


\begin{table}
\begin{tabular}{l|r@{.}lr@{.}lr@{.}lr@{.}l|r@{.}lr@{.}lr@{.}lr@{.}l}
$\bold{r}^{\sigma_s \sigma_{123}}$ & \multicolumn{8}{c|}{$\beta=7.40$} & \multicolumn{8}{c}{$\beta=7.75$}\\
\hline
$m_qa = $ & &01 & &02 & &03 & &04 & &01 & &02 & &03 & &04\\
\hline

$1+-$ & 0&2738(2) & 0&3834(2) & 0&4655(2) & 0&5332(2) & 0&2769(4) & 0&3861(4) & 0&4676(4) & 0&5351(4) \\
$1++$ & 0&93(4) & 0&98(2) & 1&02(1) & 1&063(10) & 0&540(8) & 0&617(4) & 0&687(3) & 0&749(2) \\
$3''--$ & 0&941(5) & 0&997(4) & 1&045(3) & 1&090(2) & 0&539(2) & 0&620(2) & 0&689(1) & 0&750(1) \\
$3''-+$ & 1&2(1) & 1&23(3) & 1&26(3) & 1&29(2) & 0&631(8) & 0&698(5) & 0&762(4) & 0&820(3) \\
$3''++$ & 1&18(6) & 1&25(3) & 1&31(2) & 1&36(2) & 0&67(1) & 0&731(6) & 0&793(5) & 0&849(4) \\
$3''+-$ & 1&13(1) & 1&171(7) & 1&218(5) & 1&264(4) & 0&618(4) & 0&690(2) & 0&755(2) & 0&815(2) \\
$1--$ & 1&16(3) & 1&25(2) & 1&31(1) & 1&36(1) & 0&664(7) & 0&732(4) & 0&795(3) & 0&851(3) \\

\\
\cline{1-9}

Type & \multicolumn{6}{c}{$\beta=8.00$}\\
\cline{1-9}
$m_qa = $ & &01 & &02 & &03\\
\cline{1-9}

$1+-$ & 0&2580(3) & 0&3604(3) & 0&4379(3) \\
$1++$ & 0&381(2) & 0&467(1) & 0&5393(8) \\
$3''--$ & 0&3805(7) & 0&4660(6) & 0&5379(6) \\
$3''-+$ & 0&425(2) & 0&502(1) & 0&572(1) \\
$3''++$ & 0&449(3) & 0&523(2) & 0&590(1) \\
$3''+-$ & 0&4215(10) & 0&5008(8) & 0&5698(7) \\
$1--$ & 0&447(1) & 0&523(1) & 0&5889(9) \\

\end{tabular}
\caption{Naive staggered pion masses}
\label{tab:staggered}
\end{table}


\begin{table}
\begin{tabular}{l|r@{.}lr@{.}lr@{.}lr@{.}lr@{.}l}
$\bold{r}^{\sigma_s \sigma_{123}}$ & \multicolumn{2}{c}{Asqtad} & \multicolumn{2}{c}{p4fat3} & \multicolumn{2}{c}{p4fat7} & \multicolumn{2}{c}{p4fat7tad} & \multicolumn{2}{c}{Staggered}\\
\hline

$1+-$ & 0&072(1) & 0&0675(7) & 0&1317(9) & 0&096(1) & 0&0723(4)\\
$1++$ & 0&54(1) & 0&68(3) & 0&545(9) & 0&490(9) & 0&88(4)\\
$3''--$ & 0&534(3) & 0&642(3) & 0&540(2) & 0&49(2) & 0&890(6)\\
$3''-+$ & 0&69(2) & 0&88(3) & 0&677(9) & 0&639(8) & 1&18(6)\\
$3''++$ & 0&84(3) & 1&02(10) & 0&81(2) & 0&78(2) & 1&13(5)\\
$3''+-$ & 0&719(6) & 0&841(10) & 0&691(4) & 0&658(3) & 1&07(1)\\
$1--$ & 0&83(2) & 0&93(3) & 0&802(9) & 0&77(1) & 1&11(3)\\

\end{tabular}
\caption{$m_\pi$ extrapolated to $m_q=0$ for $\beta=7.40$.}
\label{tab:chiral740}
\end{table}   


\begin{table}
\begin{tabular}{l|r@{.}lr@{.}lr@{.}lr@{.}lr@{.}l}
$\bold{r}^{\sigma_s \sigma_{123}}$ & \multicolumn{2}{c}{Asqtad} & \multicolumn{2}{c}{p4fat3} & \multicolumn{2}{c}{p4fat7} & \multicolumn{2}{c}{p4fat7tad} & \multicolumn{2}{c}{Staggered}\\
\hline

$1+-$ & 0&064(3) & 0&061(2) & 0&075(3) & 0&086(4) & 0&08(1)\\
$1++$ & 0&253(4) & 0&307(5) & 0&240(4) & 0&239(3) & 0&449(9)\\
$3''--$ & 0&257(2) & 0&314(2) & 0&244(2) & 0&243(2) & 0&449(2)\\
$3''-+$ & 0&339(5) & 0&408(6) & 0&317(4) & 0&314(4) & 0&552(9)\\
$3''++$ & 0&407(7) & 0&474(9) & 0&377(5) & 0&370(5) & 0&59(1)\\
$3''+-$ & 0&338(3) & 0&409(3) & 0&319(2) & 0&315(3) & 0&537(5)\\
$1--$ & 0&397(4) & 0&473(6) & 0&377(3) & 0&370(3) & 0&590(8)\\

\end{tabular}
\caption{$m_\pi$ extrapolated to $m_q=0$ for $\beta=7.75$.}
\label{tab:chiral775}
\end{table}    


\begin{table}
\begin{tabular}{l|r@{.}lr@{.}lr@{.}lr@{.}lr@{.}l}
$\bold{r}^{\sigma_s \sigma_{123}}$ & \multicolumn{2}{c}{Asqtad} & \multicolumn{2}{c}{p4fat3} & \multicolumn{2}{c}{p4fat7} & \multicolumn{2}{c}{p4fat7tad} & \multicolumn{2}{c}{Staggered}\\
\hline

$1+-$ & 0&046(2) & 0&050(2) & 0&058(3) & 0&052(2) & 0&063(1)\\
$1++$ & 0&152(3) & 0&191(3) & 0&142(3) & 0&138(4) & 0&268(4)\\
$3''--$ & 0&154(1) & 0&190(1) & 0&145(1) & 0&143(1) & 0&269(1)\\
$3''-+$ & 0&201(3) & 0&248(3) & 0&186(3) & 0&185(3) & 0&326(4)\\
$3''++$ & 0&237(4) & 0&290(4) & 0&219(3) & 0&222(4) & 0&357(5)\\
$3''+-$ & 0&204(1) & 0&248(1) & 0&188(1) & 0&189(1) & 0&323(2)\\
$1--$ & 0&242(2) & 0&290(2) & 0&223(1) & 0&226(2) & 0&356(2)\\

\end{tabular}
\caption{$m_\pi$ extrapolated to $m_q=0$ for $\beta=8.00$.}
\label{tab:chiral800}
\end{table}   


\begin{table}
\begin{tabular}{c|r@{.}lr@{.}lr@{.}l}
Action & \multicolumn{2}{c}{$\beta = 7.40$} & \multicolumn{2}{c}{$\beta = 7.75$} & \multicolumn{2}{c}{$\beta = 8.00$}\\
\hline

Staggered	& 1&18(2)    & 0&76(3) &  0&628(9)\\
Asqtad		& 1&09(6) & 0&73(3) &  0&602(13)\\
p4fat3		& 1&09(8) & 0&75(3) &  0&596(9)\\
p4fat7		& 0&99(3) & 0&73(3) &  0&586(6)\\
p4fat7tad	& 1&00(4) & 0&73(2) &  0&585(11)\\

\end{tabular}
\caption{$m_\rho$ extrapolated to $m_q \rightarrow 0$ for the different actions}
\label{tab:rho}
\end{table}   
\fi


\ifnum\theFigures=1
%
%






\begin{figure}
\epsfxsize=\hsize
\begin{center}
\epsfbox{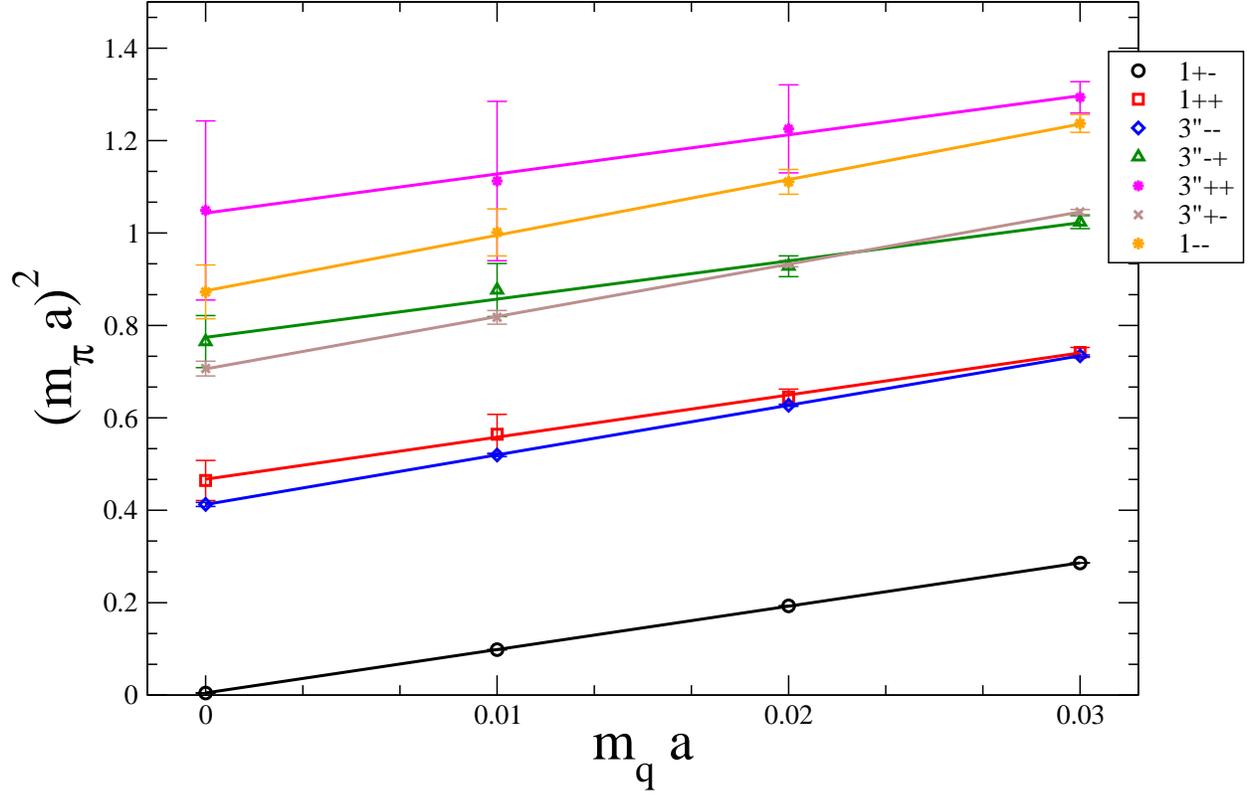}
\end{center}
\caption{Chiral extrapolation of $m_\pi^2$ for p4fat3 action, $\beta=7.40$}
\label{fig:chiral_extr}
\end{figure}

\begin{figure}
\epsfxsize=\hsize
\begin{center}
\epsfbox{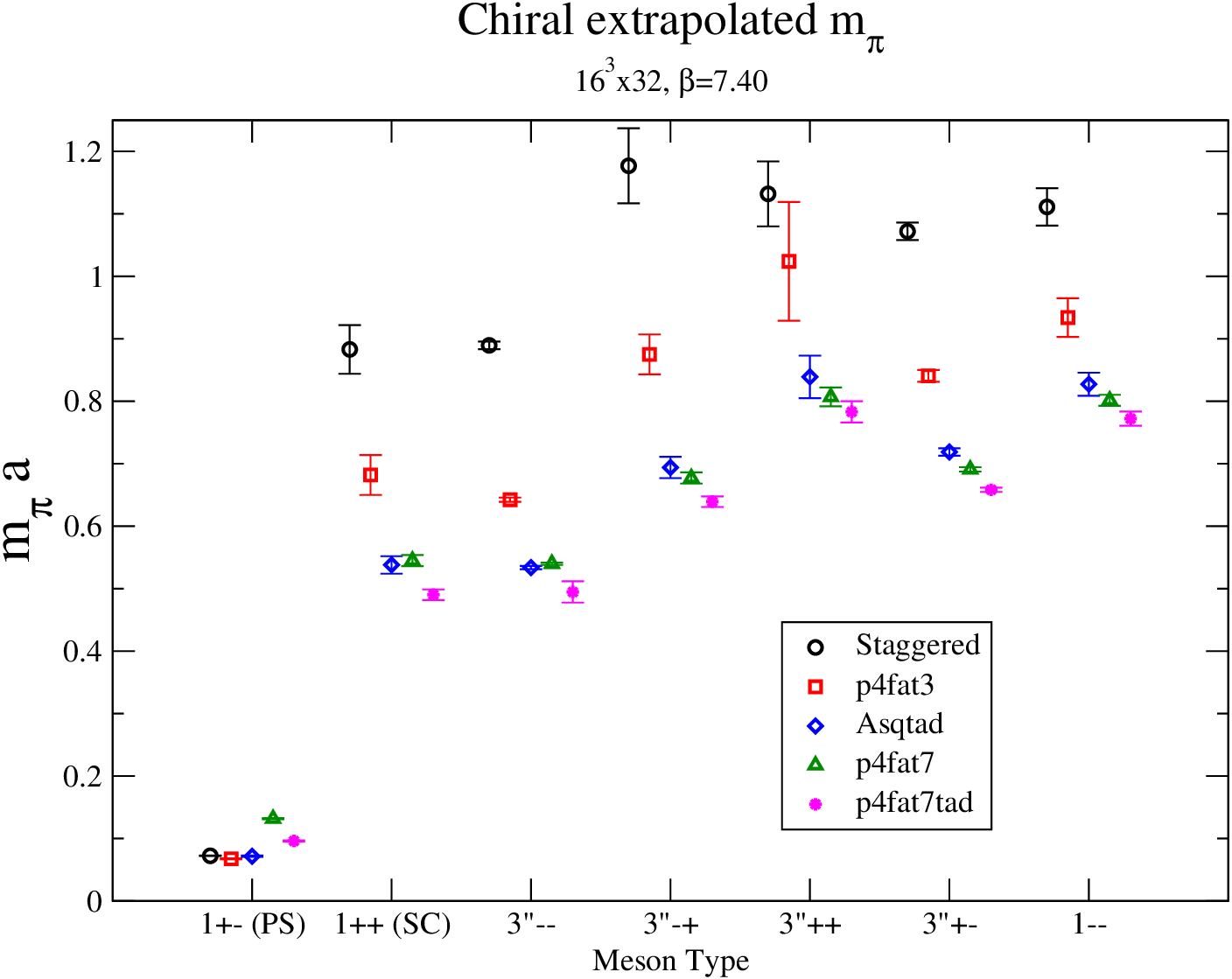}
\end{center}
\caption{Flavor symmetry breaking of $m_\pi$ in the chiral limit for $\beta = 7.40$.}
\label{fig:chiral740}
\end{figure}

\begin{figure}
\epsfxsize=\hsize
\begin{center}
\epsfbox{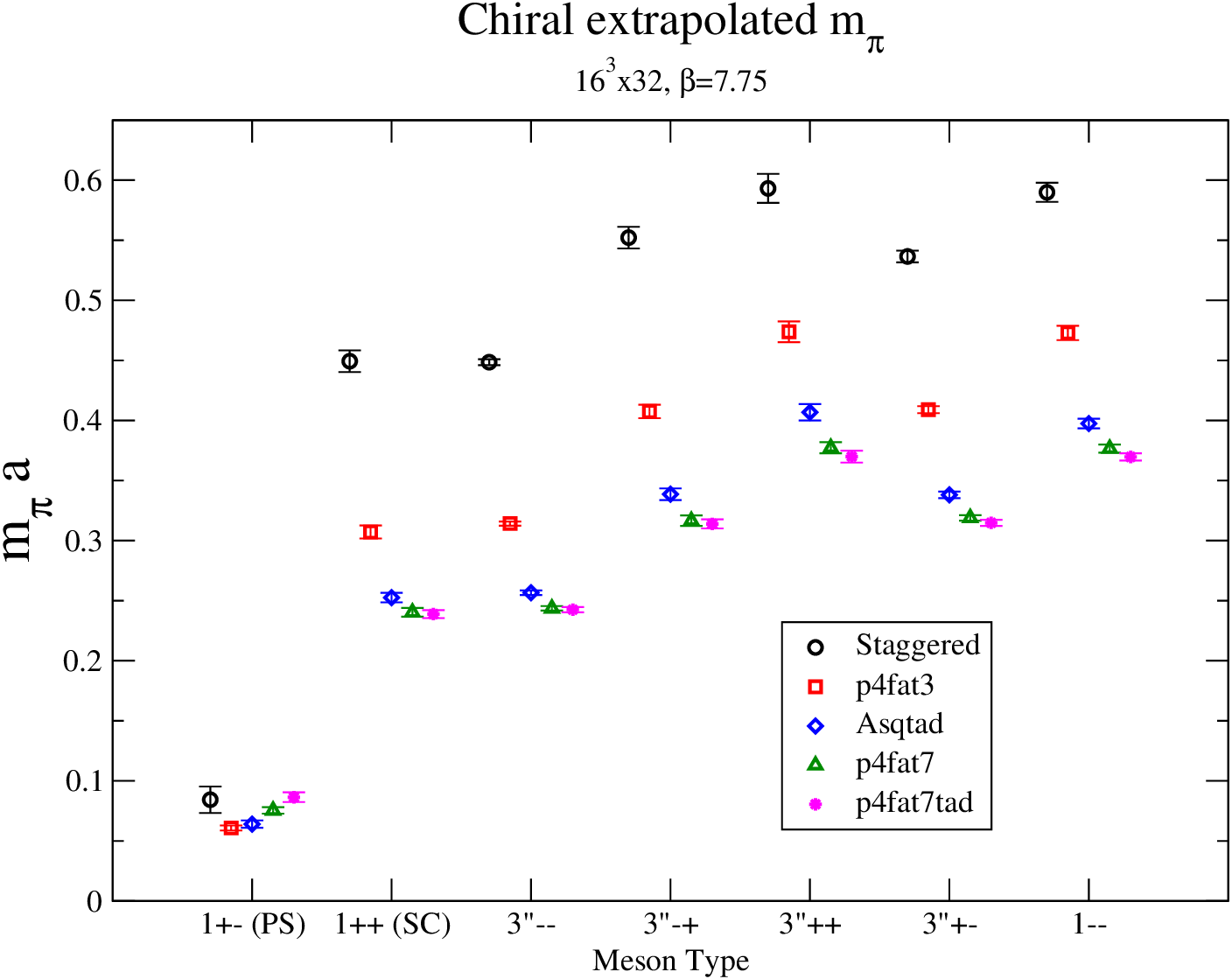}
\end{center}
\caption{Flavor symmetry breaking of $m_\pi$ in the chiral limit for $\beta = 7.75$.}
\label{fig:chiral775}
\end{figure}

\begin{figure}
\epsfxsize=\hsize
\begin{center}
\epsfbox{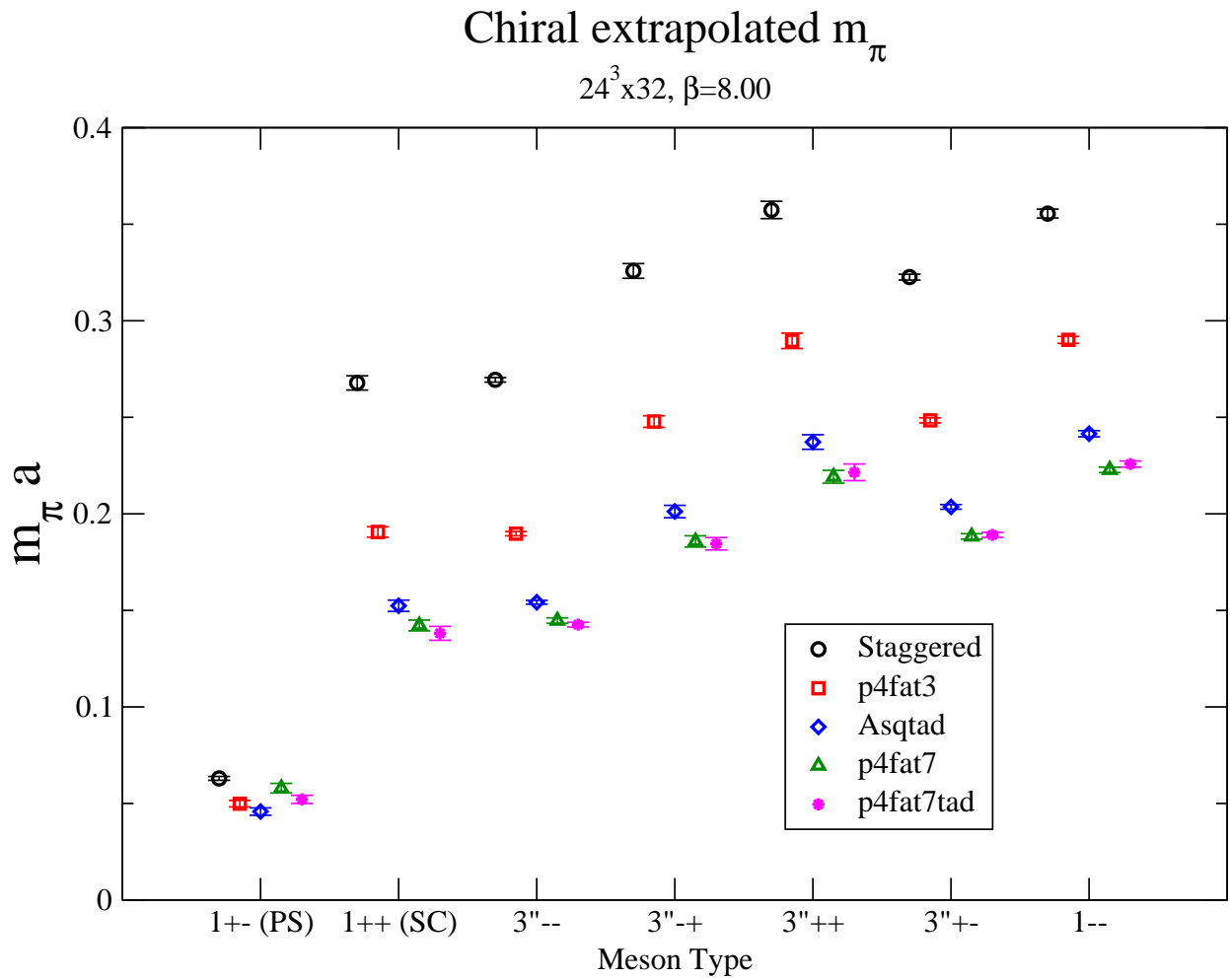}
\end{center}
\caption{Flavor symmetry breaking of $m_\pi$ in the chiral limit for $\beta = 8.00$.}
\label{fig:chiral800}
\end{figure}

\begin{figure}
\epsfxsize=\hsize
\begin{center}
\epsfbox{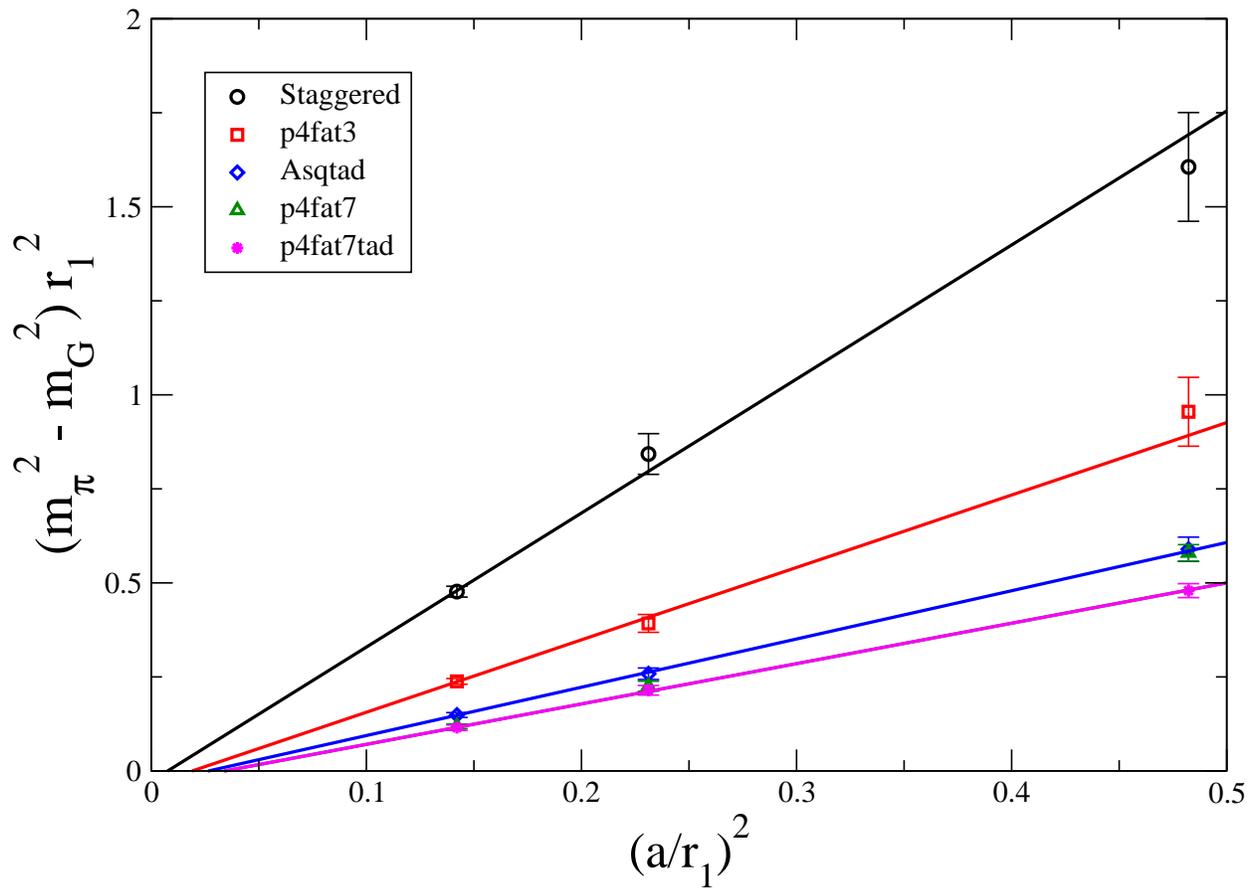}
\end{center}
\caption{Pion mass splittings for chirally extrapolated values of $m_\pi$.  $m_\pi$ corresponds to $r^{\sigma_s\sigma_{123}} = 1++$ while $m_G$ is the mass of the Goldstone pion.}
\label{fig:splitting}
\end{figure}

\begin{figure}
\epsfxsize=\hsize
\begin{center}
\epsfbox{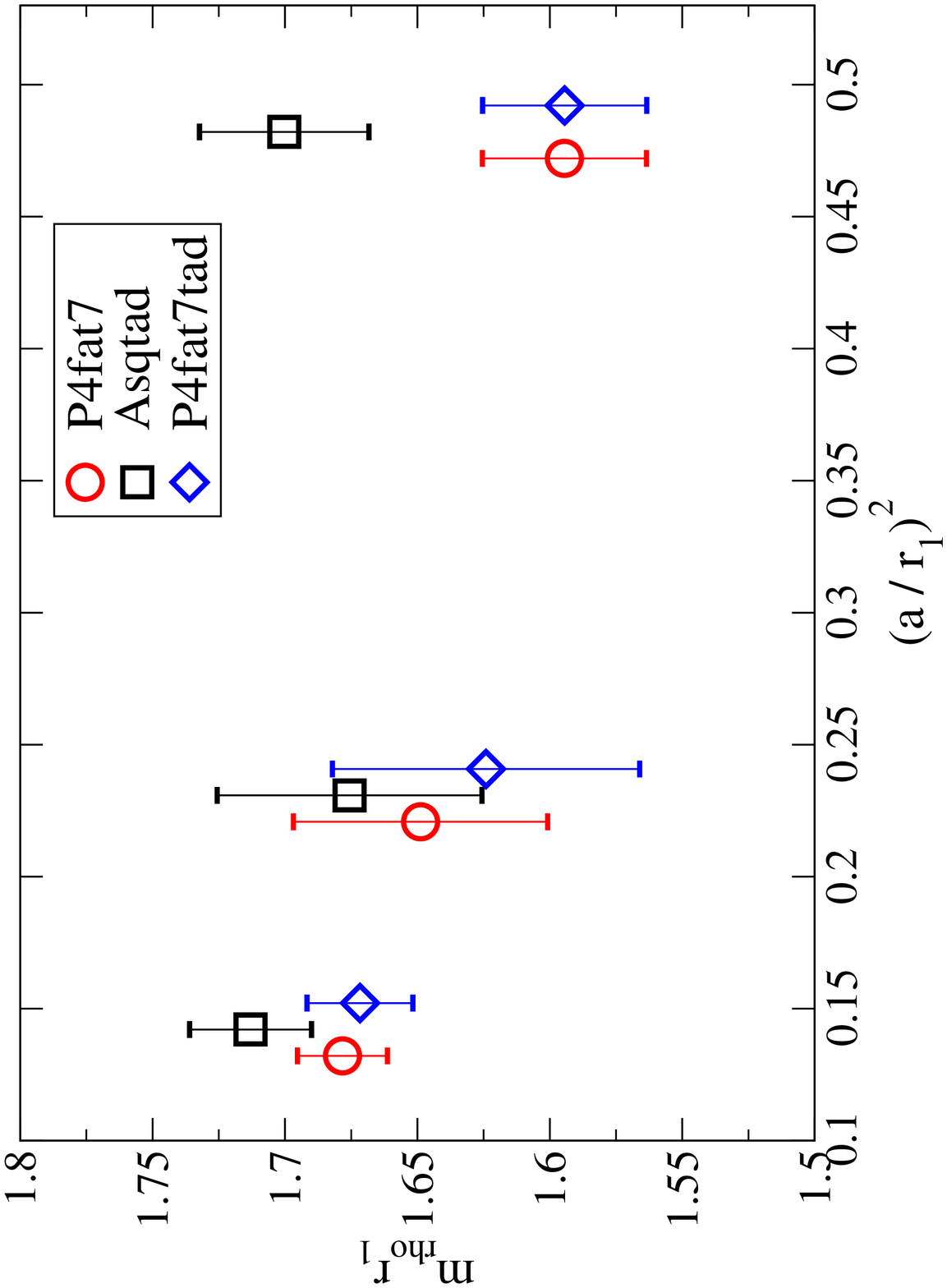}
\end{center}
\caption{Scaling of $m_\rho r_1$ vs. $(a/r_1)^2$ for various actions (points offset for clarity)}
\label{fig:rhoscaling}
\end{figure}

\fi

\end{document}